\documentclass[pra,superscriptaddress,showpacs,onecolumn,10pt]{revtex4}
\usepackage{amssymb}
\usepackage{graphicx}

\usepackage{amsthm}
\usepackage{dcolumn}
\usepackage{bm}

\def\>{\rangle}

\begin{document}
\newtheorem{question}{Question}
\newtheorem{corollary}{Corollary}
\newtheorem{definition}{Definition}
\newtheorem{example}{Example}
\newtheorem{lemma}{Lemma}
\newtheorem{proposition}{Proposition}
\newtheorem{statement}{Statement}
\newtheorem{theorem}{Theorem}
\newtheorem{property}{Property}
\newtheorem{fact}{Fact}
\newtheorem{conjecture}{Conjecture}

\newcommand{\bra}[1]{\langle #1|}
\newcommand{\ket}[1]{|#1\rangle}
\newcommand{\braket}[3]{\langle #1|#2|#3\rangle}
%%inner product
\newcommand{\ip}[2]{\langle #1|#2\rangle}
%%outer product
\newcommand{\op}[2]{|#1\rangle \langle #2|}

\newcommand{\tr}{{\rm tr}}
\newcommand {\E } {{\mathcal{E}}}
\newcommand {\F } {{\mathcal{F}}}
\newcommand {\M} {{\mathcal{M}}}
\newcommand {\f } {\tilde{F}}
\newcommand {\rr } {\tilde{r}}
\newcommand {\R } {{\mathcal{R}}}
\newcommand {\I } {{\mathcal{I}}}
\newcommand{\mod}{{\rm\  mod\ }}
\renewcommand{\b}{\mathcal{B}}
\newcommand{\h}{\mathcal{H}}
\newcommand{\T}{\mathcal{T}}

\newcommand {\Ha } {{\mathcal{H}_A}}
\newcommand {\Hb } {{\mathcal{H}_B}}
\newcommand {\diag } {{\rm diag}}

\title{Local Unambiguous Discrimination with Remaining Entanglement}
\author{Yangjia Li}
\email{liyangjia@gmail.com}
\author{Runyao Duan}
\email{Runyao.Duan@uts.edu.au}
\author{Mingsheng Ying}
\email{yingmsh@tsinghua.edu.cn}

\affiliation{State Key Laboratory of Intelligent Technology and
Systems, Tsinghua National Laboratory for Information Science and
Technology, Department of Computer Science and Technology,
Tsinghua University, Beijing 100084, China\\
 and\\
Centre for Quantum Computation and Intelligent Systems (QCIS),
Faculty of Engineering and Information Technology, University of
Technology, Sydney, NSW 2007, Australia}

\date{\today}

\begin{abstract}
A bipartite state which is secretly chosen from a finite set of
known entangled pure states cannot be immediately useful in standard
quantum information processing tasks. To effectively make use of the
entanglement contained in this unknown state, we introduce a new way
to locally manipulate the original quantum system: either identify
the state successfully or distill some pure entanglement.
Remarkably, if many copies are available, we show that any finite
set of entangled pure states, whatever orthogonal or not, can be
locally distinguished in this way, which further implies that pure
entanglement can be deterministically extracted from unknown pure
entanglement. These results make it clear why a large class of
entangled bipartite quantum operations including unitary operations
and measurements that are globally distinguishable can also be
locally distinguishable: they can generate pure entanglement
consistently.
 \end{abstract}


\maketitle

\section{Introduction and Main Results}
How to identify an unknown system which is secretely chosen from a
finite set of pre-specified states is one of the basic problems in
information theory. In quantum mechanics, it becomes more
interesting since perfect discrimination can't be achieved for
nonorthogonal states. In most cases, unambiguous discrimination is
generally effective, unless these states are linearly
dependent~\cite{Che98_DG98}. The problem is also considered when the
unknown system is shared by some physically separated parties and
only local operations on each party and classical communication
between them (LOCC) is allowed during the process~\cite{PW91}.
Things get quite different in the local discrimination. When there
are only two different states, the results are simple: the
successful probability can be always achieved optimally for both of
orthogonal case~\cite{WSHV00} and nonorthogonal case~\cite{JCY05}.
However, for more than two states, the problem becomes very
complicated. One surprising example is that some orthogonal product
states can't be perfect distinguished locally~\cite{BDF+99}. Another
one is that any orthogonal basis which is unambiguously
distinguishable by LOCC must be a product basis~\cite{HSSH03}. On
the other hand, there are also some positive results discovered such
as Ref.~\cite{BW09} and Ref.~\cite{DFJY07}

Up to now, all things considered in the above researches is how to
get classical information from the unknown system. When the process
is finished, the original system would be simply discard. Thus, if
the discrimination is failed, nothing can be obtained from it. It is
actually assumed that the only useful thing contained in this system
is distinguishability. However, this assumption doesn't conform to
the fact when people distinguish between entangled states using only
LOCC. In this scenario, the entanglement contained in the system
becomes a kind of nonclassical information, which plays
indispensable roles in standard quantum information processing tasks
such as superdense coding \cite{BW92} and
teleportation~\cite{BBC+93}. In fact, discrimination of different
states is often employed as some steps of complicated physical
tasks. Preserving as much entanglement as possible in each of these
steps will bring a lot of convenience for the whole tasks in
practice. Thus, a right choice here is to optimally make use
of the entanglement contained in the original quantum system, instead of ignoring it.

One possible strategy of doing this is to directly extract the
entanglement before discrimination and to treat it as an independent
quantum task. In this case, such a quantum task is equivalent to
distill pure entanglement from a arbitrarily given entangled mixed
state, which is known as entanglement distillation~\cite{BBP+96} in
standard quantum information processing tasks. Unfortunately, this
task is very difficult in general and most of the existing
distillation protocols which are effective for mixed states only
work in the asymptotic regime~\cite{BBP+96,BDSW96,DEJ+96,DVDV03}. It
means that the target entanglement won't become exactly pure unless
the source entanglement goes to infinite. It is still unclear how to
do this in finite regime.

What we proposed in this paper is actually another strategy which
seems more effective than the above one: we perform the entanglement
distillation as a part of our new discrimination. The idea is that
when discrimination is failed, we make the different states become
the same in the output. More important, entanglement will remain in
this outcome state while it would normally be destroyed in
traditional discrimination protocols. We noticed that some preliminary works on preserving entanglement in the local discrimination of orthogonal entangled
states have been discussed~\cite{Coh07}. However, these works are not applicable for nonorthogonal states.
Another advantage of our method is
that the original quantum system is always useful whether it is successfully
identified or not. Formally, we propose the following:

\begin{definition}\upshape
A set of bipartite pure states $\{\ket{\psi_1},\cdots,\ket{\psi_n}\}$ is said to be {\it unambiguously distinguishable with remaining entanglement}(UDRE) via LOCC if there is a local measurement $\{M_s,M_f\}$ such that i) $\{M_s\ket{\psi_k}\}$ are locally distinguishable with certainty; ii) $M_f\ket{\psi_k}\equiv\ket{\eta}$ for some bipartite entangled state $\ket{\eta}$, where the subscripts ``$s$" and ``$f$" represent the successful and the failed outcomes of discrimination, respectively. Note that we write $\ket{\alpha}\equiv \ket{\beta}$ if there is scalar $\lambda$ such that $\ket{\alpha}=\lambda\ket{\beta}$.
\end{definition}

Clearly, any two orthogonal pure states $\ket{\psi_0}$ and
$\ket{\psi_1}$ are locally UDRE \cite{WSHV00}. However, if nonorthogonal or
more than two states are involved, the situation becomes quite
complicated. For instance, it is not difficult to show that any two
nonorthogonal or three orthogonal $2\otimes 2$ pure entangled states
cannot be distinguishable in this way. It is also easy to see that
two nonorthogonal pure states containing at least one product state
cannot be UDRE (Actually from the remaining entangled state we can
infer that the original state should be entangled and thus achieve a
perfect discrimination, which is a contradiction with the
nonorthogonality).

For $2\otimes 3$ states there do exist nonorthogonal pure states
which can be distinguished via UDRE. An explicit example is given as
follows:
\begin{eqnarray}
\ket{\psi_0}&=&(\ket{0}_A\ket{0}_B+\ket{1}_A\ket{1}_B+\ket{0}_A\ket{2}_B)/\sqrt{3},\nonumber\\
\ket{\psi_1}&=&(\ket{0}_A\ket{0}_B+\ket{1}_A\ket{1}_B+\ket{1}_A\ket{2}_B)/\sqrt{3}.\nonumber
\end{eqnarray}
A simple protocol is as follows: Bob performs a measurement
$\{\op{0}{0}+\op{1}{1},\op{2}{2}\}$. If the outcome is $2$ then
Alice only needs to perform the measurement
$\{\op{0}{0},\op{1}{1}\}$ to complete the discrimination. Otherwise
the discrimination is failed and they are left with an entangled
state $(\ket{00}+\ket{11})/\sqrt{2}$.

It is still unknown when a set of states can be locally UDRE. Interestingly, if multiple copies are available, we find
that any finite number of bipartite pure entangled states can be
distinguished in this way.

\begin{theorem}\label{the:UDRE}\upshape
Let $\ket{\psi_1},\cdots,\ket{\psi_n}$ be any $n$ bipartite
entangled pure states in $\Ha\otimes\Hb$. Then there is always an
integer $N$ such that $\ket{\psi_1}^{\otimes
N},\cdots,\ket{\psi_n}^{\otimes N}$ are UDRE.
\end{theorem}

As a direct consequence, we have the following

\begin{theorem}\label{the:main_distillation}\upshape
Let $\ket{\psi_1},\cdots,\ket{\psi_n}$ be any $n$ bipartite
entangled pure states on $\Ha\otimes\Hb$. Then there is always an
integer $N$ and an LOCC protocol $\mathcal{E}$ such that
$\mathcal{E}(\psi_k^{\otimes
N})=\op{\Phi_0}{\Phi_0}$ for any $1\leq k\leq N$, where $\ket{\Phi_0}=(\ket{0}\ket{0}+\ket{1}\ket{1})/\sqrt{2}$.
\end{theorem}

The above theorem has a clear physical meaning: one can use LOCC to
deterministically distill an Einstein-Podolsky-Rosen (EPR)
state~\cite{EPR35} of form $\ket{\Phi_0}$ from the following of
mixed state
$$\rho(N)=\sum_{i=1}^{n}p_i[\op{\psi_i}{\psi_i}]^{\otimes N},$$
where $p_i\geq 0$ and $\sum_{i=1}^n p_i=1$. For $n=1$, it is a
special case of the fundamental result in entanglement
transformation~\cite{Nie99}. However, our generalization here brings
it to a essentially different level since the source entanglement
becomes mixed. This distillation protocol for mixed states works in
finite regime and succeed with probability one. Compared with
traditional distillation protocols, it has two remarkable
advantages: first, both of the amount of source entanglement and the
number of runs of LOCC can be bounded before execution; second,
without statistical fluctuation, the result is more faithful and
reliable. Although only states of form $\rho(N)$ is given in
our distillation protocol, it works for all UDRE states.

\section{Proof of theorem~\ref{the:UDRE}}

We will now provide a complete proof to Theorem~\ref{the:UDRE}. To
make our arguments more readable, we first consider the case of
$n=2$ and then generalize the result to the case where $n>2$.

Suppose the dimensions of $\Ha$ and
$\Hb$ are $d'$ and $d$, respectively, and $d'\geq{d}\geq{2}$. Let us assume one of
$\ket{\psi_1}$ and $\ket{\psi_2}$ is of full Schmidt number at Bob's side. Later in Appendix $1$ we will show this assumption can be removed. To be specific, suppose that $\ket{\psi_2}$ is with the Schmidt decomposition as follows:
\begin{equation}\label{equ:Sch-phi}
\ket{\psi_2}=\sum_{i=0}^{d-1}\sqrt{\lambda_{i}}\ket{i_A i_B},~\mbox{where}~\lambda_i>0, 0\leq i\leq d-1.
\end{equation}
Let us emphasize that $\{\ket{i}_B: 0\leq i\leq d-1\}$ is an orthonormal basis of $\Hb$ and $\{\ket{i}_A: 0\leq i\leq d-1\}$ is a set of orthonormal vectors of $\Ha$. Let
$\lambda_{\min}=\min\{\lambda_i:0\leq i\leq d-1\}.$  A key property of  full Schmidt rank state is the following:

\begin{lemma}\label{observation}\upshape
Let $\ket{\psi_1}$ and $\ket{\psi_2}$ be two bipartite pure states on $\Ha\otimes \Hb$ such that $\ket{\psi_2}$ is of full Schmidt rank at Bob's side. Then there is a linear operator $M_A$ such that
$(M_A\otimes I_B)\ket{\psi_2}=\ket{\psi_1}$ and $M_A^\dagger M_A\leq \lambda_{\min}^{-1} I_A$.
\end{lemma}

{\bf Proof.} Note that $\ket{\psi_1}$ can be uniquely decomposed as
$\ket{\psi_1}=\sum_{i=0}^{d-1}\ket{\alpha_i}\ket{i}_B,$
where $\{\ket{i}_B:0\leq i\leq d-1\}$ is the same as in Eq. (\ref{equ:Sch-phi}) and $\{\ket{\alpha_i}:0\leq i\leq d-1\}$ is a set of (unnormalized) vectors on $\Ha$. Let $M_A=\sum_{i=0}^{d-1}\lambda_i^{-1/2}\op{\alpha_i}{i_A}$. One can readily verify that $M$ satisfies our requirements.\hfill $\blacksquare$\\

The following lemma provides a sufficient condition for UDRE.
\begin{lemma}\label{lem:special_case}\upshape
Let $\ket{\alpha}$ and $\ket{\beta}$ be two states
on an auxiliary system $\mathcal{H}_{A'}$ of Alice. Then $\ket{\alpha}_{A'}\ket{\psi_1}_{AB}$ and
$\ket{\beta}_{A'}\ket{\psi_2}_{AB}$ are locally unambiguously distinguishable with
remaining entanglement $\ket{\alpha}_{A'}\ket{\psi_1}_{AB}$ if
$|\ip{\alpha}{\beta}|\leq\sqrt{\lambda_{\min}/(1+\lambda_{\min})}.$
\end{lemma}

{\bf Proof.}
Let us write $\ket{\alpha}=\ket{0}$ and $\ket{\beta}=u\ket{0}+v\ket{1}$, where
$u=\ip{\alpha}{\beta}$. By a simple algebraic calculation we have
$|u/v|\leq \sqrt{\lambda_{\min}}$.
We are now seeking for a linear operator $O_{A'A}$ such that i)\ ${O}^\dagger O\leq I_{A'A}$; ii) $O\ket{0}_{A'}\ket{i}_A=\ket{0}_{A'}\ket{i}_A/\sqrt{2}, ~0\leq i\leq d-1$; and iii) $(O_{A'A}\otimes
I_B)\ket{1}_{A'}\ket{\psi_2}=1/\sqrt{2}(u/v)\ket{0}_{A'}(2\ip{\psi_1}{\psi_2}\ket{\psi_1}-\ket{\psi_2})$.

If such a linear operator $O_{A'A}$ exists, we can construct a local measurement $\{M_s,
M_f\}$ such that $M_f=O_{A'A}\otimes I_B$ and $M_s=\sqrt{I_{A'A}-O^\dagger O}\otimes I_B$. It is quite straightforward to verify that such a local measurement can achieve a UDRE.

Consider the following linear operator
$$O=1/\sqrt{2}\op{0}{0}_{A'}\otimes I_A+1/\sqrt{2}\op{0}{1}_{A'}\otimes (u/v) M_A,$$
where $M$ is a linear operator such that $(M_A\otimes I_B)\ket{\psi_2}=(2\ip{\psi_1}{\psi_2}\ket{\psi_1}-\ket{\psi_2})$ and $M_A M_A^\dagger \leq \lambda_{\min}^{-1}I_A$. The existence of $M$ follows directly from Lemma \ref{observation}. Clearly, by construction conditions ii) and iii) are automatically satisfied. The validity of condition i) follows from the observation  $OO^\dagger=1/2\op{0}{0}\otimes I_A+1/2\op{0}{0}\otimes (u/v)^2 M M^\dagger \leq \op{0}{0}\otimes I_A.$ \hfill $\blacksquare$\\

What is missing in Lemma \ref{lem:special_case} is how to construct
the auxiliary system $A'$ used on the Alice's part when the given
state is $\rho(N)$. Our idea is to transform the state
$\ket{\psi_1}^{\otimes N_0}$ (resp. $\ket{\psi_2}^{\otimes N_0}$) to
$\ket{\alpha}$ (resp. $\ket{\beta}$) for some $N_0$, and then system
$A'$ can be constructed from these $N_0$ copies. The following lemma plays a key role.
\begin{lemma}\label{lem:decomposition}\upshape
There exists an orthonormal basis $\{\ket{e_{i}}: 0\leq i\leq d-1\}$ of $\mathcal{H}_B$ such that
\begin{eqnarray}
\ket{\psi_1}_{AB}&=&\sum_{i=0}^{d-1}a_{i}\ket{\alpha_{i}}_A\ket{e_{i}}_B,~\ip{\alpha_i}{\alpha_i}=1,\nonumber\\
\ket{\psi_2}_{AB}&=&\sum_{i=0}^{d-1}b_{i}\ket{\beta_{i}}_A\ket{e_{i}}_B,~\ip{\beta_i}{\beta_i}=1,\nonumber
\end{eqnarray}
and $|\ip{\alpha_{i}}{\beta_{i}}|<1$ for each $0\leq i\leq d-1$. In particular, $(\ket{\psi_1}_{AB},\ket{\psi_2}_{AB})$ can be transformed into $(\ket{\alpha}_A,\ket{\beta}_A)$ by a local operation, where $\ip{\alpha}{\beta}=\max\{|\ip{\alpha_{i}}{\beta_{i}}|:0\leq i\leq d-1\}< 1$.
\end{lemma}

{\bf Proof.} For the case of $d=2$, we can choose
$\{\ket{e_{0}},\ket{e_{1}}\}$ to be one of the following three bases:
$\{\ket{0},\ \ket{1}\}$, $\{\ket{+},\ \ket{-}\}$, and
$\{\frac{4}{5}\ket{0}+\frac{3}{5}\ket{1},\
\frac{3}{5}\ket{0}-\frac{4}{5}\ket{1}\}$.

Now we start to consider the general case $d>2$. Let us start with the basis $\ket{e_i}=\ket{i}_B$ which gives the Schmidt decomposition of $\ket{\psi_2}$. That is, $\ip{\alpha_i}{\alpha_j}=\delta_{ij}$. Let $I_1=\{0\leq i\leq d-1: \ket{\alpha_i}\equiv\ket{\beta_i}\}$. If $I_1=\emptyset$ then we have done. Otherwise we have the following two cases:

Case 1. $I_1=\{0,\cdots, d-1\}$. In this case $\ket{\psi_1}$ and $\ket{\psi_2}$ are simultaneously Schmidt decomposable. Let us define a new basis $\ket{e_i'}_B=1/\sqrt{d}\sum_{j=0}^{d-1} \omega^{ij}\ket{e_j}_B$, where $\omega=e^{2\pi i/d}$ is the $d$-th root of unity. Then we have
$\ket{\alpha_i'}_A=\sum_{i=0}^{d-1}\omega^{ij} a_i \ket{\alpha_i}_A$
and
$\ket{\beta_i'}_A=\sum_{i=0}^{d-1}\omega^{ij} b_i \ket{\beta_i}_A.$
One can readily verify that $\ip{\alpha_i'}{\beta_i'}=\ip{\psi_1}{\psi_2}$. That completes our proof in this case.

Case 2. $I_1\subsetneqq\{0,\cdots, d-1\}$. To be specific, assume that $\ket{\alpha_0}\equiv \ket{\beta_0}$ and $\ket{\alpha_1}\not\equiv\ket{\beta_1}$. Then consider the following two sub-vectors of $\ket{\psi_1}$ and $\ket{\psi_2}$:
\begin{eqnarray}
\ket{\psi_1'}&=&a_0\ket{\alpha_0}_A \ket{e_0}_B+a_1\ket{\alpha_1}_A\ket{e_1}_B,\nonumber\\
\ket{\psi_2'}&=&b_0\ket{\beta_0}_A\ket{e_0}_B+b_1\ket{\beta_1}_A\ket{e_1}_B.\nonumber
\end{eqnarray}
Applying the result for $d=2$ we know that there is an orthonormal basis $\{\ket{e_0'}_B,\ket{e_1'}_B\}$ of  $span\{\ket{e_0}_B, \ket{e_1}_B\}$ such that $\ip{e_0'}{\psi_1'}\not \equiv \ip{e_0'}{\psi_2'}$ and $\ip{e_1'}{\psi_1'}\not \equiv \ip{e_1'}{\psi_2'}$. Thus under the new basis $\{\ket{e_0'}_B,\ket{e_1'}_B, \ket{e_2}_B,\cdots, \ket{e_{d-1}}_B\}$, we have $I_1'=I_1-\{0\}\subsetneqq I_1$.  Repeating this process at most $d-1$ times we can finally have $I_1=\emptyset$.\hfill $\blacksquare$\\

For $\ket{\alpha}$ and $\ket{\beta}$ in the above lemma, we can choose a positive integer $N_0$ such that
$$|\ip{\alpha}{\beta}|^{N_0}\leq \sqrt{\frac{\lambda_{\min}}{1+\lambda_{\min}}}.$$
Then it follows from Lemma \ref{lem:special_case} that
\begin{corollary}\label{cor:N02_pure}\upshape
$\ket{\psi_1}^{\otimes(N_0+1)}$ and $\ket{\psi_2}^{\otimes(N_0+2)}$
are locally unambiguously distinguishable with remaining entangled
state $\ket{\psi_2}$.
\end{corollary}

This completes the proof of Theorem \ref{the:UDRE} for the case
where $k=2$ and one of $\ket{\psi_1}$ and $\ket{\psi_2}$ has full
Schmidt number. The proof for the most general case is somewhat
involved and is given in Appendix $1$.

\section{Applications to the Local Identification of Quantum Measurements}

We will employ Theorem \ref{the:main_distillation} to study the local identification of quantum operations, which formalize all physically realizable operations in quantum mechanics~\cite{NC00}. Recently the problem of distinguishing quantum operations has attracted a lot of attentions. Many interesting results have been reported~\cite{A01_DLP01,JFDY06,WY06_Wat07_PW09,DFY07,DFY09}. It was
shown that perfect identification can be achieved for unitary
operations~\cite{A01_DLP01} and projective
measurements~\cite{JFDY06}. A necessary and sufficient condition for the perfect distinguishability of quantum operations has been discovered quite recently~\cite{DFY09}. An important generalization of this problem is to consider the identification of a bipartite unknown quantum operation using LOCC only, which seems much more complicated than the global setting. Surprisingly, it has been shown that the perfect discrimination between unitary operations is always possible even under LOCC \cite{DFY08}. However, the local distinguishability of general quantum operations remains unknown so far.

Here we consider the identification of bipartite quantum measurements.
We will employ a new strategy different from that in \cite{DFY08}, i.e., we will reduce the LOCC discrimination problem to the global case by generating pure bipartite entanglement using the known apparatus. Applying our result about deterministic distillation, we can generate a sufficiently large number of EPR pairs between Alice and Bob, and thus accomplish the local perfect discrimination by teleportation and global protocol. This motivates us to introduce the following
\begin{definition}\label{def:consistent_entanglers}\upshape
Two quantum measurements $\mathcal{M}_0=(M_{01},\cdots, M_{0n_0})$ and $\mathcal{M}_1=(M_{11},\cdots, M_{1n_1})$ acting on
$\Ha\otimes\Hb$ are said to be \emph{consistently entangled}, if there
exists $\ket{\alpha}\in\mathcal{H}_{A'}\otimes \Ha$ and $\ket{\beta}\in \mathcal{H}_{B'}\otimes \Hb$ such that
$(I_{A'}\otimes I_{B'}\otimes M_{ik})\ket{\alpha}_{A'A}\ket{\beta}_{B'B}$ is entangled or vanished for $1\leq i\leq n_k$ and $k=0,1$, where $A'$ and $B'$ are auxiliary systems.
\end{definition}

It follows immediately from Theorem \ref{the:main_distillation} that bipartite pure entanglement can be extracted by a finite number of runs of an unknown operation $\mathcal{M}$ which is secretely chosen from
two consistently entangled measurements $\mathcal{M}_0$ and $\mathcal{M}_1$. So if $\mathcal{M}_0$ and
$\mathcal{M}_1$ are perfectly distinguishable by global quantum operations, then they are also perfectly distinguishable by LOCC.

If we focus our attention on the identification of two bipartite projective
measurements $\mathcal{M}_0=\sum_{i=1}^l{iP_i}$ and $\mathcal{M}_1=\sum_{j=1}^l{jQ_j}$ and apply the fact that any two projective measurements are identifiable by global operations~\cite{JFDY06}, we have the following results about the local distinguishability of consistently entangled projective measurements.
\begin{lemma}\label{cor:consistent_entanglers_case}\upshape
If $\mathcal{M}_0$ and $\mathcal{M}_1$ are consistently entangled bipartite projective measurements, then they are perfectly distinguishably by LOCC.
\end{lemma}

Unfortunately, for $\mathcal{M}_0$ and $\mathcal{M}_1$ that may not be consistently entangled, the local distinguishability remains unknown. Nevertheless, we still have the following sufficient condition. The proof of this result is somewhat tricky and is put in Appendix $2$.
\begin{lemma}\label{lem:generalization_M_U_M}\upshape
Two projective measurements $\mathcal{M}_0$ and $\mathcal{M}_1$ can be perfectly distinguished by LOCC if for some $k\in\{1,2,\cdots,l\}$ there is a product state in the supports of $P_k$ or $Q_k$ but not in their intersection.
\end{lemma}

To describe the above two identifiable cases in a more unified way, we will employ the notion of Unextendible Product Bases (UPB)~\cite{BDM+99_DMS+03}. A UPB is a set
of orthogonal product pure states which cannot be further extended
by adding any additional orthogonal product state. The notion of UPB is very
important as it can be used to construct bound entanglement or to demonstrate the weird phenomenon ``quantum nonlocality without entanglement" ~\cite{DFJY07}. It can also be used to construct interesting examples in quantum information theory ~\cite{SST01_DY07}. In addition, we need to introduce the notion of
\emph{Unextendible Product Part} (UPP). Let
$W$ be a set of product states $\ket{\alpha\beta}$ such that $\ket{\alpha\beta}$ is in the intersection of the supports of $P_k$ and $Q_k$ for some $k$.  That is,
$$W=\{\ket{\alpha\beta}:\ket{\alpha\beta}\in supp(P_k)\cap supp(Q_k)~\mbox{for some }k\}.$$
A subset $X$ of $W$ is said to be a (orthogonal) \emph{Product Part} (PP) of $\mathcal{M}_0$ and $\mathcal{M}_1$,
if any two states in $X$ are orthogonal. A PP is called a UPP if
it can be a proper subset of any other PP.
\begin{theorem}\upshape
$\mathcal{M}_0$ and $\mathcal{M}_1$ are perfectly distinguishable if they have a UPP which is not a UPB.
\end{theorem}

{\bf Proof.} Let $Y$ be such a UPP. Notice that $|Y|<dim(\mathcal{H}_A\otimes\mathcal{H}_B)$ as $\mathcal{M}_0$ and $\mathcal{M}_1$ are
different. Since $Y$ is not a UPB, there is a product state $\ket{\alpha\beta}$ in
the orthogonal complement of $Y$. If for every $k$, both
$P_k\ket{\alpha\beta}$ and $Q_k\ket{\alpha\beta}$ are not product
states, then we can identify $\mathcal{M}$ according to Corollary
\ref{cor:consistent_entanglers_case}. Otherwise, without loss of
generality, we can assume that
$P_k\ket{\alpha\beta}=\lambda\ket{\gamma\eta}(\lambda\neq0)$ is a
product state. Then $\ket{\gamma\eta}$ is in the support of $P_k$ as
$P_k\ket{\gamma\eta}=P_kP_k\ket{\alpha\beta}=P_k\ket{\alpha\beta}=\ket{\gamma\eta}$.
We claim that $\ket{\gamma\eta}$ is not in the support of $Q_k$, and
then the identifiability of $\mathcal{M}$ follows immediately from Lemma
\ref{lem:generalization_M_U_M}. In fact, if it is not the case, then
$\ket{\gamma\eta}$ is in the both supports of $P_k$ and $Q_k$. Thus, $Y\cup\{\ket{\gamma\eta}\}$ is a PP which strictly includes $Y$. This contradicts our assumption that $Y$ is a
UPP. \hfill $\blacksquare$\\

If the dimension of one part of a bipartite system is $2$, then any UPP of $\mathcal{M}_0$ and $\mathcal{M}_1$ cannot be a UPB as there is no UPB for $2\otimes n$ quantum system \cite{BDM+99_DMS+03}. So we have:
\begin{theorem}\upshape
Any two projective measurements acting on $2\otimes n$ are perfectly distinguishable by LOCC.
\end{theorem}

However, there does exists two locally indistinguishable bipartite projective measurements if the dimension of each subsystem is not less than $3$. This is essentially due to the existence of UPB for these quantum systems \cite{BDM+99_DMS+03}.
Let $\{\ket{\alpha_i}_A\ket{\beta_i}_B\}_{i=1}^l$ be a UPB and
$P=\sum_{i=1}^l\op{\alpha_i\beta_i}{\alpha_i\beta_i}$. Partition
$I_{AB}-P$ into two nontrivial orthogonal projectors $Q_1$ and $Q_2$. Then we claim that the
following two bipartite projective measurements
\begin{eqnarray}
\mathcal{M}_0&=&\sum_{i=1}^li\op{\alpha_i\beta_i}{\alpha_i\beta_i}+(l+1)(I-P),\nonumber\\
\mathcal{M}_1&=&\sum_{i=1}^li\op{\alpha_i\beta_i}{\alpha_i\beta_i}+(l+1)Q_1+(l+2)Q_2\nonumber
\end{eqnarray}
cannot be perfectly distinguishable using LOCC. Actually, due to the property of UPB, $\mathcal{M}_0$ and $\mathcal{M}_1$ will yield an outcome $r\in \{1,\cdots, l\}$ with nonzero probability on any product input state. This makes further perfect discrimination impossible as both output states are $\ket{\alpha_r\beta_r}$.

\section{Conclusion}
In this paper, we have considered the problem of how to effectively manipulate a quantum system whose state is
secretely chosen from a set of nonorthogonal bipartite pure entangled states.
Our idea is to distill some pure entanglement when the discrimination is failed.
Then the notion of {\it Unambiguous Discrimination with Remaining
Entanglement} has been introduced to describe such a discrimination
protocol. We have shown that an unknown state which
belongs to a finite set of bipartite pure entangled states $\{\ket{\psi_i}^{\otimes
N}\}$ can be locally distinguished via UDRE for some sufficiently large $N$.
This discrimination protocol is then used to construct exact and
deterministic entanglement distillation protocols. Most interestingly, a
mixed state $\rho(N)$ produced by randomly chosen from a finite set
of $N$-copy entangled states $\{\ket{\psi_i}^{\otimes N}\}$, can be
locally transformed to an EPR pair $\ket{\Phi_0}$, for some
sufficiently large $N$.

We apply our distillation to identify an unknown bipartite quantum
measurement locally. It provides us a way to obtain EPR pairs
by a finite number of runs of this unknown measurement. The local
identification has been reduced to global identification in these
cases. In particular, we have discussed the local identification from two
bipartite projective measurements and have found a computable
sufficient condition of identifiable measurements. Surprisingly,
even in this simple case, we have found an example of two bipartite
measurements which are locally undistinguishable. It exposes the
difference between global identification and local identification,
for quantum measurements.

We are grateful to Nengkun Yu for suggesting a different approach to
deal with the general case of Theorem \ref{the:UDRE}. We also thank Yuan Feng and Zhengfeng Ji
for useful discussions. This work was
partially supported by NSF of China (Grant Nos. 60621062, 60702080,
60736011) and QCIS, University of Technology, Sydney.

\section*{Appendix 1: Proof of the general case of Theorem \ref{the:UDRE}}

In this Appendix we turn to prove Theorem \ref{the:UDRE} for the
general case $\{\ket{\psi_1},\ket{\psi_2},\cdots,\ket{\psi_n}\}$.
Let us first complete the proof of $n=2$ case. we will show that the
full Schmidt rank assumption can be removed. In fact, we will show
that for any two bipartite entangled pure states $\ket{\psi_1}$ and
$\ket{\psi_2}$, by performing a local projective measurement
$\{P_0', P_1'\}$ on Bob's side we can either i) with outcome ``0"
achieve a successful discrimination between these two states, or ii)
with outcome ``1" obtain another two entangled states
$\ket{\psi_1'}$ and $\ket{\psi_2'}$ on a smaller state space
$\mathcal{H}_{A'}\otimes \mathcal{H}_{B'}$, and one of them is of
full Schmidt rank.

Let $\Hb(\psi_1)$ and $\Hb(\psi_2)$) denote the supports of $\tr_A(\op{\psi_1}{\psi_1})$
and $\tr_A(\op{\psi_2}{\psi_2})$), respectively. Let $P_1$ and $P_2$ be
their respective projectors. We have $rank(P_1),\
rank(P_2)\geq 2$ as $\ket{\psi_1}$ and $\ket{\psi_2}$ are both
entangled. The case of $\ip{\psi_2}{\psi_1}= 0$ is trivial. In
the following, we assume that $\ip{\psi_2}{\psi_1}\neq 0$. Obviously,
$P_2P_1\neq 0$. Now we complete the proof for $n=2$ by considering the
following three cases:
\begin{itemize}
\item Case 1. $rank(P_2P_1)\geq 2$.  It suffices to choose a measurement $\{P_0'=I_B-P_2, P_1'=P_2\}$ on Bob's
part. Outcome ``0" indicates the original state is $\ket{\psi_1}$
while outcome ``1" yields a pair of entangled states on
$\Ha\otimes\Hb(\phi_2)$, namely $I_A\otimes P_2\ket{\phi_1}$ and the
full Schmidt number state $I_A\otimes P_2\ket{\phi_2}=\ket{\phi_2}$,
and the latter case has been proven.

\item Case 2. $rank(P_2P_1)=1$, and $P_1P_2=P_2P_1$. Let
$P_2P_1=\op{x}{x}$, where $\ket{x}$ and $\ket{y}$ are both
normalized states in $\Hb$. $P_1\ket{x}=\ket{x}$,
$P_2\ket{x}=\ket{x}$, $Q_1:=P_1-\op{x}{x}\perp P_2$, and
$Q_2:=P_2-\op{y}{y}\perp P_1$, where $Q_1$ and $Q_2$ are projectors
of rank at least $1$. Let $\ket{1}$ and $\ket{2}$ be eigenstates of
$Q_1$ and $Q_2$, respectively. Consider a measurement
$\{P_0'=I_B-P_1', P_1'=\op{x}{x}+\op{1}{1}+\op{2}{2}\}$. It is easy
to verify that outcome ``0" results in two orthogonal states, and
outcome ``1" results in two new bipartite entangle pure states with
projectors $P_1=\op{1}{1}+\op{x}{x}$ and $P_2=\op{2}{2}+\op{x}{x}$,
respectively.  Let $\ket{z_\pm}=(\ket{1}\pm\ket{2})/\sqrt{2}$. By
another local measurement $\{M_1=\op{z_+}{z_+}+\op{x}{x}/\sqrt{2},
M_2=\op{z_-}{z_-}+\op{x}{x}/\sqrt{2}\}$ we finally obtain two
entangled states with the same supports $\op{z_+}{z_+}+\op{x}{x}$ or
$\op{z_-}{z_-}+\op{x}{x}$, which is again a proven case.

\item Case 3: $rank(P_2P_1)=1$ and  $P_2P_1\neq P_1P_2$. Let $P_2P_1=a\op{y}{x}$, where $\ket{x}$ and $\ket{y}$ are both
normalized states in $\Hb$ and $0<|a|<1$. It is easy to check that
$\ip{x}{y}\neq 0$, $P_1\ket{x}=\ket{x}$, $P_2\ket{y}=\ket{y}$,
$Q_1=P_1-\op{x}{x}\perp P_2$, and $Q_2=P_2-\op{y}{y}\perp P_1$,
where $Q_1$ and $Q_2$ are projectors of rank at least $1$.  Bob can
perform a local measurement $\{P_0'=I_B-P_1',
P_1'=Q_1+Q_2+\op{x}{x}\}$. Then he can either determine the original
state is $\ket{\psi_1}$ or yield another pair of entangled states
$\ket{\psi_1'}=(I_A\otimes P_1')\ket{\psi_1}=\ket{\psi_1}$ and
$\ket{\psi_2'}=(I_A\otimes P_1')\ket{\psi_2}$, which is reduced to
Case 2 as $\ket{x}$ is in both supports.
\end{itemize}

Now we consider the case of $n>2$. We shall first prove the result
for the following three special states on two pairs of qubits:
$\ket{\psi_1}_{AB}=\ket{\Phi_0}_{{A_1}{B_1}}\ket{\Phi_0}_{{A_2}{B_2}}$,
$\ket{\psi_2}_{AB}=\ket{\phi_1}_{{A_1}{B_1}}\ket{\Phi_0}_{{A_2}{B_2}}$,
$\ket{\psi_3}_{AB}=\ket{\Phi_0}_{{A_1}{B_1}}\ket{\phi_2}_{{A_2}{B_2}}$.
If $\ket{\phi_1}$ is an entangled state, $\{\ket{\Phi_0},
\ket{\phi_1}\}$ is a proven case on pair $1$, and the result is
reduced to $n=2$ case. So it suffices to consider that
$\ket{\phi_1}$ is a product state. Similarly, we assume
$\ket{\phi_2}$ to be a product state in the same way. Without loss
of generality, let
$\ket{\phi_1}=(a\ket{0}+b\ket{1})_{A_1}\ket{0}_{B_1}$ and
$\ket{\phi_2}=(c\ket{0}+d\ket{1})_{A_2}\ket{0}_{B_2}$, where
$|a|^2+|b|^2$=1 and $|c|^2+|d|^2=1$.  (The reason that we can choose
$\ket{\phi_1}$ (or $\ket{\phi_2}$) of this form is that we can
perform $U_A\otimes {U_A}^*$ on pair $1$ (pair $2$) without change
the state $\ket{\Phi_0)}$ on both pairs).

Now let us introduce a linear operator $O_{B_1B_2}$ on Bob's system $B_1B_2$ as follows:
$$O_{{B_1}{B_2}}=\ket{00}(ac\bra{00}+ad\bra{01}+bc\bra{10})+\ket{11}\bra{11}.$$
Then $\mathcal{M}=\{M_0, M_1\}$ is a local measurement, where
$M_1=I_{A_1A_2}\otimes O_{B_1B_2},
M_0=I_{A_1A_2}\otimes\sqrt{I_{B_1B_2}-O^\dagger{O}}$. Notice that
$\braket{\psi_2}{M_0^\dagger M_0}{\psi_3}=0$. So after this
measurement, $\ket{\psi_2}$ and $\ket{\psi_3}$ become orthogonal if
the outcome is ``0". and then, we can reduce the present case to
$n=2$ case by distinguishing between these two states. Thus, we only
need to consider that when the outcome is always ``1". In this
setting, it's easy to check that $M_1\ket{\psi_1}$ is an entangled
state $\ket{\xi}$ while $M_1\ket{\psi_2}, M_2\ket{\psi_3}$ become
the same state $\ket{\psi}=\ket{\alpha}_{A_1A_2}\ket{00}_{B_1B_2}$.
On the other hand, by the results in $n=2$ case of Theorem
\ref{the:main_distillation}, $\ket{\psi_2}^{\otimes N}$ and
$\ket{\psi_3}^{\otimes N}$ can be transformed to $\ket{\Phi_0}$ at
the same time by some local measurement $\mathcal{M}'$. Now, measure
$\{\ket{\psi_1}^{\otimes (N+1)}$, $\ket{\psi_2}^{\otimes (N+1)}$,
$\ket{\psi_3}^{\otimes (N+1)}\}$ by $\mathcal{M}$ on the first copy
and $\mathcal{M}'$ on the other $N$ copies. Suppose $\mathcal{M}'$
brings $\ket{\psi_1}^{\otimes N}$ into some $\ket{\phi}$. Then the
whole state becomes a mixture of two entangled states
$\ket{\xi}_{{A_1}{B_1}}\ket{\phi}_{{A_2}{B_2}}$ and
$\ket{\psi}_{{A_1}{B_1}}\ket{\Phi_0}_{{A_2}{B_2}}$, which is a
proven case.

Finally, let us complete the whole proof by induction on $n\geq 2$.
By induction hypothesis and the result in Theorem
\ref{the:main_distillation}, let local measurements $\mathcal{M}$
and $\mathcal{M}'$ transform $\ket{\psi_1}^{\otimes N_1}$,
$\ket{\psi_2}^{\otimes N_1}$, $\cdots$, $\ket{\psi_{n-1}}^{\otimes
N_1}$ and $\ket{\psi_2}^{\otimes N_2}$, $\ket{\psi_2}^{\otimes
N_2}$, $\cdots$, $\ket{\psi_n}^{\otimes N_2}$ to $\ket{\Phi_0}$,
respectively. Thus, $\ket{\psi_1}^{\otimes (N_1+N_2)}$,
$\ket{\psi_2}^{\otimes (N_1+N_2)}$, $\cdots$,
$\ket{\psi_{n}}^{\otimes (N_1+N_2)}$ can be reduced to a proven
case. In fact, measure the first $N_1$ copies of this unknown state
by $\mathcal{M}$ and measure the rest $N_2$ copies by
$\mathcal{M}'$. Suppose $M$ takes $\ket{\psi_n}^{\otimes N_1}$ to
some $\ket{\phi_1}$ and $\mathcal{M}'$ takes $\ket{\psi_1}^{\otimes
N_2}$ to some $\ket{\phi_2}$. Then these states become
$\ket{\Phi_0}\ket{\phi_2}, \ket{\Phi_0}\ket{\Phi_0},
\ket{\phi_1}\ket{\Phi_0}$ for $\ket{\psi_1}, \ket{\psi_i}(i\neq 1,
k), \ket{\psi_k}$ respectively. This is exactly the case we have
just proven for $n=3$.\hfill $\blacksquare$

\section*{Appendix 2: Proof of Lemma \ref{lem:generalization_M_U_M}}

Without loss of generality, we may assume that $\ket{\alpha\beta}$ is in the support of $Q_k$ only. So we have
$Q_k\ket{\alpha\beta}=\ket{\alpha\beta}$ and $P_k\ket{\alpha\beta}=\lambda\ket{\Phi}$, where
$\lambda\neq{0}$ (Otherwise $\mathcal{M}$ can be identified according the
outcome of the measurement on $\ket{\alpha\beta}$) and
$|\ip{\alpha\beta}{\Phi}|<1$. First, Alice and Bob apply $\mathcal{M}$ to input state $\ket{\alpha}\ket{\beta}$. If the outcome is not $k$, then $\mathcal{M}=\mathcal{M}_0$ and the
identification is finished. Otherwise, the output state is
$\ket{\Phi}$ and $\ket{\alpha\beta}$. To finish the proof, we need to consider the following two cases separately:

Case 1. $\ket{\Phi}=\ket{\gamma}_A\ket{\delta}_B$
is a product state. Clearly we have $|\ip{\Phi}{\alpha\beta}|<1$. If we further $|\ip{\Phi}{\alpha\beta}|\leq1/4$, then $|\ip{\alpha}{\gamma}|\leq1/2$ or
$|\ip{\beta}{\delta}|\leq1/2$. Without loss of generality we may assume that  $|\ip{\gamma}{\alpha}|\leq1/2$, then
it follows from \cite{identification_M} there is a local isometry $U$ acting on
$\mathcal{H}_A$ satisfying  $U\ket{\gamma}\perp\ket{\gamma}$ and $U\ket{\alpha}=\ket{\alpha}$. A perfect discrimination can be achieved by applying $(U\otimes I)$ to $(\ket{\Phi}, \ket{\alpha\beta})$ and then applying $\mathcal{M}$ one more time. If $|\ip{\alpha\beta}{\Phi}|>1/4$, we can apply $\mathcal{M}$ in parallel to $n$ copies of $\ket{\alpha\beta}$ yielding two output product states with inner product $|\ip{\alpha\beta}{\Phi}|^n\leq 1/4$ for sufficiently large $n$.

Case 2. $\ket{\Phi}$ is entangled. Employing Lemma \ref{lem:decomposition}, we can transform $(\ket{\Phi_{AB}},\ket{\alpha_A\beta_B})$ locally into $(\ket{\gamma_A0_B},\ket{\alpha_A0_B})$ for some $\ket{\gamma}$ such that $|\ip{\gamma}{\alpha}|<1$. Similarly we can transform $(\ket{\Phi_{AB}},\ket{\alpha_A\beta_B})$ locally into $(\ket{0_A\delta_B},\ket{0_A\beta_B})$ for some $\ket{\delta}$ such that $|\ip{\delta}{\beta}|<1$. Combining these two transformations we can achieve the following transformation locally: $(\ket{\Phi_{AB}}^{\otimes 2}, \ket{\alpha_A\beta_B}^{\otimes 2})$ to $(\ket{\gamma_A\delta_B},\ket{\alpha_A\beta_B})$. It is clear by repeating this process sufficiently large times we can obtain two pairs of product states $\ket{\gamma_A\delta_B}$ and $\ket{\alpha_A\beta_B}$ such that $|\ip{\gamma}{\alpha}|$ and $|\ip{\delta}{\gamma}|$ can be arbitrarily small but not zero. Now we are trying to show that by choosing $\ket{\gamma_A\delta_B}$ carefully we can have $P_k\ket{\gamma\delta}\perp Q_k\ket{\alpha\beta}$, or equivalently, $P_k\ket{\alpha\beta}\perp \ket{\gamma\delta}$. This is obvious as the orthogonal complement of $P_k\ket{\alpha\beta}$ is spanned by a set of product states, and not all of these states are orthogonal to $\ket{\alpha\beta}$. Otherwise $P_k\ket{\alpha\beta}$ should coincide with $\ket{\alpha\beta}$, a contradiction with our assumption.\hfill $\blacksquare$
\end{document}